\documentclass[a4paper]{article}
\usepackage[left=2.00cm, right=2.00cm, top=2.5cm, bottom=2.500cm]{geometry}

\usepackage[utf8]{inputenc}
\usepackage[T1]{fontenc}
\usepackage[english]{babel}
\usepackage{amsmath}
\usepackage{amsfonts}
\usepackage{amssymb}
\usepackage{graphicx}
\usepackage{authblk}

\begin{document}
\title{Trend detection in GEV models}
\title{Trend detection in GEV models\thanks{CONTACT L. N\'emeth. Email: lnemeth@caesar.elte.hu}%\thanks{Grants or other notes
}


\author[1]{L\'aszl\'o N\'emeth}
\author[2]{Zuzana H\"ubnerov\'a}
\author[3]{Andr\'as Zempl\'eni}

\affil[1,3]{E\"otv\"os Lor\'and University,  Institute of Mathematics, Department of Probability Theory and Statistics, Budapest, Hungary}
\affil[2]{Brno University of Technology, Institute of Mathematics, Brno, Czech Republic}

\maketitle

\begin{abstract}
	In recent environmental studies extreme events have a great impact. The yearly and monthly maxima of environment related indices can be analysed by the tools of extreme value theory. For instance, the monthly maxima of the fire weather index in British Columbian forests might be modelled by GEV distribution, but the stationarity of the time series is questionable. This property can lead us to different approaches to test if there is a significant trend in past few years data or not. An approach is a likelihood ratio based procedure which has favourable asymptotic properties, but for realistic sample sizes it might have a large error. In this paper we analyse the properties of the likelihood ratio test for extremes by bootstrap simulations and aim to determine a minimal required sample size. With the theoretical results we re-asses the trends of fire weather index in British Columbian forests.
\end{abstract}

% 60G70: Extreme value theory
% 62M10: Time series, auto-correlation, regression

\section{Introduction}

Extreme value theory propose numerous methods to analyse the distribution of extreme events. Fisher-Tippet theorem describes the limit distribution of block maxima for iid. observations, peaks over threshold models introduce the Pareto limit distribution. 
However, in real life identical distribution can not be guaranteed in every cases, it is possible that trends are present in the data (e.g. \cite{clarke}, \cite{ribei}). In this paper we propose trend detection methods for extreme value data, motivated by a real life environmental problem.

Estimating the fire hazard in forests is an important application of environmental statistics. The numerous risk factors (such as temperature, wind, draught etc.) can be converted into a single number, that is called fire weather index (FWI) \cite{fwidata}. The FWI is used in Canada for measuring fire potential and it can change day by day and also has a seasonality. In the past years, multiple data were collected at stations in British Columbia and the daily FWI is available between $1970-2015$. However, global warming and other environmental factors can affect the fire potential. Therefore, it might be of interest to identify stations with temporal changes in FWI.

In a previous analysis, Hrdli\v{c}kov\'{a}, Esterby and Taylor aimed at
clustering the fire weather stations according to their typical trend during the years \cite{TIES}. Also, recently, a relationship of FWI and climate change was of the interest of this research group \cite{chapter} using separate \cite{coles} as well as spatial max-stable models \cite{davison}. In the analysis, they studied the trends in monthly maximal FWI in every year. By the Fisher-Tippet theorem \cite{ft}, the distribution of the maximal FWI can be modelled by a generalized extreme value (GEV) distribution undel mild conditions if the sample is independent and identically distributed. However, it is suspected, that there is a linear trend in the monthly maxima of FWI. For detecting a trend,
two different approaches are available for analysing the data as mentioned in \cite{clarke}. First, we can fit a GEV model with trend in location $\mu$ and scale $\sigma$ parameter, instead of iid. GEV model, see \cite{coles} or \cite{panag}. Second, as in \cite{clarke}, we may fit a linear regression to the data and suppose the residuals as GEV distributed. To identify the significance of the trend one may use likelihood ratio test in each model. Accomplishing goodness of fit test on the sample and on the residuals of the best fitting linear model it can be chosen which station satisfies the theoretical property (i.e. monthly maxima of FWI follow a GEV distribution). These likelihood ratio tests also provide p-values to measure the significance of a linear trend (assuming that the sample size is large enough for the asymptotic properties to hold). We compare the two approaches by analytical results and by working on real datasets and propose an alternative to the usual likelihood ratio test. 

In our analysis, we used bootstrap methods, introduced by Efron \cite{efron}, to understand the structure of the data and to construct confidence intervals for the estimations. Besides classical bootstrap and permutation bootstrap for extreme values \cite{permboot}, we have also used parametric bootstrap \cite{kyseley}. Resampling methods can be used for construction confidence intervals for trend coefficient under different models. Additionally we present analytically calculated return level estimations which are confirmed by simulations.

Unfortunately, the data collection periods at different stations were not homogeneous (at some stations data collection started later, at others it stopped for periods of different length), thus many of the maxima are missing. The small sample size and the GEV-like error can cause falsely discovered trend in many cases. In our analysis, we aim to find a minimal required sample size, which allows for using the likelihood ratio test to identify a linear trend of given size at given power and Type 1 error. This could help to decide which station's data is acceptable for analysis and in which case more years  of observations or different methods for detecting trend are needed. Additionally, we analyse the dependence of minimal required sample size on the estimated tail index parameter by simulations and give a small-sample alternative to the usual likelihood-ratio based inference. 

In our analysis we intend to find the sufficient sample size for the considered LR tests to have acceptable type $1$ error and the highest possible power. In \cite{cai} the normality of the estimate of the GEV return levels "as a criterium for sufficient sample size" is investigated.
There are also methods available for estimating the shape parameter for samples of relatively small sizes \cite{regress}.
Some small simulation studies similar to our analysis can be found in \cite{hasonlo}. Unlike the likelihood ratio test in our analysis, paper \cite{panag} focuses on AIC and BIC criteria for choosing between stationary and nonstationary models. They also investigate the uncertainity in estimation by confidence intervals. Other methods for constructing the confidence intervals for GEV parameters are given in \cite{dupuis}.

The paper is organised as follows. In Section \ref{methods} we describe trend detections methods for extreme value data. First we introduce the likelihood ratio test for the significance of the slope in the linear trend of the location parameter. It is studied by bootstrap methods and parametric simulations in Section \ref{GEV_paratrend}. Section \ref{section_LM} presents the results of the second approach based on subtracting the linear trend from the original time series estimated by least squares or Theil-Sen method and handling the residuals as iid. GEV distributed.

In Section \ref{simulation} we apply parametric and non-parametric bootstrap simulations to approximate the type 1 error of the likelihood ratio tests as well as their power. We compare the effectiveness of methods for simulated samples.

Finally, in Section \ref{real} we analyse the FWI dataset, using the presented methods. We calculate the significance of the slope, which can be assessed by its confidence interval. In section \ref{CI_section} we present the confidence intervals obtained by the bootstrap method. Additionally, we calculate $5$ years return levels for FWI.

\subsection{About the data}
	
The dataset, we were able to use for testing contains monthly FWI maxima between $1970-2015$, observed at $15$ stations in British Columbia (the most representative ones were chosen from a bigger database). In most stations, the number of available observations is between $10$ and $20$. For insufficient sample size we needed to exclude $2$ of them, the further analysis includes only $13$. Before the analysis, we examine the distribution of the monthly maxima to decide if a GEV model is acceptable or not. We applied Cramér-von Mises and Anderson-Darling tests to analyse goodness of fit. 
%For stations $365,791$, we experienced significant difference from GEV distribution (possible trend), but for stations $19,68,75,105,145,161,363,401,427,447,865$ the extreme value model without any trend is acceptable at the significance level 0.05.

\section{Methods}\label{methods}

\subsection{GEV distribution with trend in parameters}\label{GEV_paratrend}

Our first aim was to check if there was a significant trend in the time series of maxima, using the likelihood ratio method, by fitting GEV distribution to the data in each station. Primarily, we were interested in identifying a linear trend in the location parameter that is feasible in two following ways. Let us call the model of iid. GEV distributed sample M0, the GEV model with trend in location (scale) parameter M1.mu (M1.sigma) and the GEV model with trend in both location and scale parameters M2. Comparing M0 and M1.mu models or M1.sigma and M2 by the likelihood ratio test could detect a significant trend in location parameter. The two approaches differ in fixing the scale parameter or accepting  a possible trend in the scale parameter as well. 
More precisely, the considered likelihood ratio statistics are
\begin{eqnarray}
LR_1({\mathbf X})&=& 2 ( {\, \ell_{GEV}(\hat{\mu}_0+\hat{\mu}_1 {\mathbf t},\hat{\sigma},\hat{\xi},{\mathbf X} )} - 
{\, \ell_{GEV}(\tilde{\mu},\tilde{\sigma},\tilde{\xi},{\mathbf X} )}) \\
LR_2({\mathbf X})&=& 2( {\, \ell_{GEV}(\hat{\mu}_0+\hat{\mu}_1 {\mathbf t},\hat{\sigma}_0 +  \hat{\sigma}_1 {\mathbf t},\hat{\xi},{\mathbf X} )} -
{\,\ell_{GEV}(\tilde{\mu},\tilde{\sigma}_0 +  \tilde{\sigma}_1 {\mathbf t},\tilde{\xi},{\mathbf X} )} )
\end{eqnarray}
where the tilde above a parameter denotes the maximum likelihood estimate (MLE) of the parameter under the null hypothesis, which is in model M1.scale (for $LR_2$) and model M0 (for $LR_1$), and the hat above a parameter denotes the maximum likelihood estimate of the parameter under alternative, which is in model M2 (for $LR_2$) and model M1.mu (for $LR_1$). Also $\ell_{GEV}(\mu,\sigma,\xi,\mathbf X)$ stands for loglikelihood of GEV with location parameter $\mu$, scale parameter $\sigma$, shape parameter $\xi$ under a sample of $\mathbf X$. In both cases, the LR statistics should have asymptotically $\chi^2(1)$ distribution under the null hypothesis - under regularity conditions that hold for the GEV for $\xi$ >-0.5 cases. Note, that hereinafter we  refer to the parameter $\mu_1$ as trend. And a test of a trend means a test of hypothesis H$_0: \mu_1 = 0$.

\subsection{Linear trend with GEV residuals}\label{section_LM}
The second studied method for analysing a trend is fitting a linear regression model by least squares (we refer this model as M3) or another robust method, called Theil-Sen estimator (model M4). Due to the nature of the data, we suppose the residuals of the linear model as iid. GEV distributed. This model is simpler than the model in Section \ref{GEV_paratrend}, however it can not handle possible heteroscedasticity, which can be problematic in the real applications. Nonetheless, we saw that in most of the cases the homoscedasticity could be accepted.

After fitting a linear regression we investigated the residuals of the model. Our assumption is that the distribution of the residuals is GEV, thus we estimate the GEV parameters. It is worth mentioning, that under linear regression model the residuals are usually not independent \cite{clarke}, which might affect the ML estimator of the GEV parameters of the residuals.
In our analysis we overcome this problem with bootstrap simulations \cite{freedman} and proposing modifications on likelihood ratio test.
%Q: earlier - bevezetésbe: mi ezt fogjuk csinálni. Új alfejezet az elejére, LR-ről.
After estimating the GEV parameters of the residuals, we could compare our model with the best fitting iid. GEV distribution on the data. This test could decide if the regression model is more supported by the data than the iid. GEV.
The likelihood ratio test statistics can be expressed as follows
\begin{eqnarray}
LR_3({\mathbf X})&=& 2 ( {\,\ell_{GEV}({\mu}_0^{LS}+{\mu}^{LS}_1 {\mathbf t}+\hat{\mu},\hat{\sigma},\hat{\xi},{\mathbf X})} -
{\,\ell_{GEV}(\tilde{\mu},\tilde{\sigma},\tilde{\xi},{\mathbf X} )}) \\
LR_4({\mathbf X})&=& 2({\,\ell_{GEV}({\mu}_0^{TS}+{\mu}^{TS}_1 {\mathbf t}+\hat{\mu},\hat{\sigma},\hat{\xi},{\mathbf X})} -
{\,\ell_{GEV}(\tilde{\mu},\tilde{\sigma},\tilde{\xi},{\mathbf X} )} )
\end{eqnarray}
where $\mathbf{t}$ stands for time, $LS$ and $TS$ stands for least squares estimate and Theil-Sen estimate, respectively. Also the hat denotes the MLE estimates under the assumption of iid. GEV distribution of the residuals, whereas tilde denotes the MLE estimated under the assumption of iid. GEV of the original vectors. Since the GEV distribution belongs to the location-scale family \cite{lehmann}, the alternative hypothesis can be formulated under one GEV distribution, where we get the location parameter as adding the estimated value of linear trend and the location parameter of the GEV fitted to the residuals (${\mu}_0+{\mathbf t}{\mu}_1+\hat{\mu}$), while the other parameters originated also from the fitted GEV.

In the first attempt, we used ordinary least squares method for fitting the linear trend. However, we experienced that extreme values behave as outliers, thus the estimation could be biased in some cases. For solving this problem, we used the more robust Theil-Sen estimation \cite{theilsen,wilcox} of trends.
Let $(x_1,y_1),(x_2,y_2),\dots,(x_l,y_l)$ be the sample, where $x$ is the independent and $y$ is the dependent variable. Calculate $$T(i,j)=\frac{y_j-y_i}{x_j-x_i}$$ for every $1\le i< j \le l$. Then the Theil-Sen estimation for the trend is the $median({T(i,j), 1\leq i<j\leq l})$ value. It is not sensitive to extremely high or small values, moreover under classical assumptions (i.e. normally distributed residuals) it behaves similarly to the least square method \cite{theilsen}.

Likelihood ratio test statistics is only asymptotically $\chi^2$ distributed. Our analysis in Section \ref{type1} shows, that the $LR_1$ and $LR_2$ test for extremes usually have more than $0.05$ type 1 error probability for small ($10-80$) samples, while $LR_3$ and $LR_4$ test have much less. To equalize these properties and make the methods generally acceptable we simulated critical values, with which the likelihood ratio tests have approximately $0.05$ type 1 error probability for certain GEV distributed data. We will refer this procedure as modified likelihood ratio tests.

\section{Simulation study}\label{simulation}

In this section we analyse the properties of LR tests on simulated samples, with similar behaviour to our real life data. We approximate the type 1 error using iid. GEV samples, we present power calculations on simulated samples containing trend.

\subsection{Type 1 error}\label{type1}
\subsubsection{$LR_1$ and $LR_2$}
We were interested in the effectiveness of likelihood ratio test on theoretically iid. GEV distributed data with no trend, to approximate the type $1$ error of $LR_1$ and $LR_2$ tests. For the analysis, we simulated varying size samples from GEV distribution with different tail index ($\xi$) parameters (see Table \ref{table:type1}). The location ($\mu$) and scale ($\sigma$) were set as the average of fitted model parameters to the real data (i.e. $\mu=22,\sigma=10$). We declared significant the cases, where $\mu$ had a significant trend using likelihood ratio test at the significance level 0.05. As one can see, the ratio of false discovery (type $1$ error) decreases for larger sized samples, moreover higher tail index causes slightly more error. This phenomena can be explained by the properties of distributions with high tail index - it is likely that there is an extreme value in the sample, which functions as an outlier and generates a false trend. As a result, we can say, that for realistic tail indices ($-0.5<\xi<0.5$, see the results in Table \ref{table:stations}) at least $30-50$ observations are required to reach the $0.05$ type $1$ error. 

The analysis was executed by using the {\tt fevd()} and {\tt lr.test()} functions from package {\tt extRemes} in R programming language. The {\tt fevd()} function uses L-moments and Gumbel moments for initializing the parameters, however in a non-negligible number of cases the fitting method diverges, which results in inappropriate parameter estimations. We experienced such a problem on samples from a GEV distribution with small scale and negative shape parameter. Thus, we used the parameters of the maximum likelihood estimates, assuming iid. GEV model ({\tt gev.fit()} from package {\tt ismev}) as initial estimates of the parameters in the {\tt fevd()} iteration.

\subsubsection{$LR_3$ and $LR_4$}
For calculating type 1 error for the regression model we applied simulation based analysis. We simulated independent samples from GEV$(22,10,\xi)$ distribution of size 10-80. Based on $1000$ simulations we calculated the false discovery ratio of the $LR_3$ test at the significance level 0.05, i.e. the empirical type 1 error of the test. The results in Table \ref{table:type1} show, that in most of the cases $20$ sample elements are sufficient to reach about $0.1$ as the probability of type 1 error, regardless of the shape parameter.

Since the least square method can not handle outliers well, we also considered the $LR_4$ test, which is based on Theil-Sen estimator as in \cite{clarke}. 
For normally distributed samples the least square and Theil-Sen estimators are similar, however for distributions with heavier tail, like GEV with high shape parameter, the two methods can be different. The results in Table \ref{table:type1} show, that the type 1 error of $LR_4$ test is closer to the desired $0.05$, especially for high $(\xi=0.5)$ tail index.

\subsubsection{Comparison}
\begin{figure}[h!]
	\centering
	\includegraphics[width=0.6\textwidth]{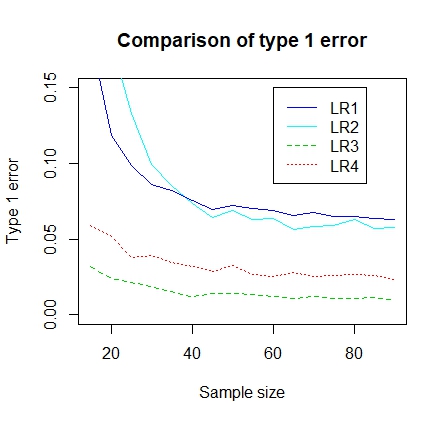}
	\caption{Comparison of type 1 error of $LR_1$,$LR_2$ $LR_3$ and $LR_4$ tests based on $10000$ simulations. Data was simulated from $GEV(22,10,0.5)$ distribution of sizes between $15$ and $90$. }
	\label{fig:type1}
\end{figure}

In Figure \ref{fig:type1} one can compare the type $1$ error of $LR_1$, $LR_2$, $LR_3$ and $LR_4$ tests on $GEV(22,10,0.5)$ data. The simulation shows that the type $1$ error of $LR_3$ is much less than $0.05$, which may results in worse power. In that aspect, $LR_4$ is closer to the ideal $0.05$. Note the case of $\xi=0.5$, where the difference between the simulated and theoretical type 1 error is the most remarkable. Nevertheless, similar results with smaller differences can be seen in other cases too.

\begin{table}[h]
	\centering
	\caption{Simulated type 1 error of likelihood ratio test of the trend under $LR_1$, $LR_2$, $LR_3$ and $LR_4$ calculated by $1000$ simulations from $GEV(22,10,\xi)$ distribution with varying shape parameters and different sample sizes.}
	\label{table:type1}
	\begin{tabular}{||c|c|c|c|c|c|c|c|c|c|c|c|c||   }
		\hline\hline
		&\multicolumn{3}{|c|}{$LR_1$}&\multicolumn{3}{|c|}{$LR_2$}&\multicolumn{3}{|c|}{$LR_3$}&\multicolumn{3}{|c|}{$LR_4$}\\
		Sample size: & 20&40&80& 20&40&80& 20&40&80& 20&40&80 \\
		\hline\hline
		$\xi=-0.5$&0.221&0.102&0.082&0.108&0.054&0.059&0.129&0.054&0.053&0.098&0.052&0.045\\
		\hline
		$\xi=-0.25$&0.16&0.086&0.056&0.127&0.061&0.057&0.081&0.054&0.055&0.077&0.049&0.042\\
		\hline
		$\xi=0$&0110&0.066&0.062&0.142&0.049&0.048&0.068&0.063&0.045&0.069&0.053&0.042\\
		\hline
		$\xi=0.25$&0094&0.05&0.062&0.163&0.063&0.051&0.066&0.049&0.034&0.074&0.05&0.035\\
		\hline
		$\xi=0.5$&0.12&0.069&0.068&0.173&0.09&0.055&0.032&0.02&0.012&0.051&0.035&0.025\\		
		\hline\hline
		
	\end{tabular}
\end{table}

\subsection{Statistical power}

\subsubsection{Used tests}
An additional analysis was performed to assess the statistical power of the likelihood ratio tests by parametric bootstrap. Under a specific alternative we have a trend in the parameters of the GEV sample, thus classical bootstrap (sampling with replacement) is not adequate in this case.
It would be worth testing all the methods, as misspecification might occur. 
Therefore we use parametric bootstrap. By \cite{coles}, every GEV distributed sample can be transformed to a standard Gumbel distribution. 
If $X_1,X_2,\dots,X_k\sim GEV(\mu,\sigma,\xi)$ then 
\begin{equation}
	Y_1,Y_2,\dots,Y_k\sim GEV(0,1,0)\sim Gumbel(0,1)
	\label{stgumbel}
\end{equation}
 where $Y_i=\log(1+\xi\cdot(X_i-\mu)/\sigma)/\xi$. This transformation can be applied even if the data is not iid, but we know the distribution of each observation. Resampling from the transformed values or simulating iid. sample from $Gumbel(0,1)$ gives the bootstrap sample after the inverse transformation.

For power calculation, we examined a GEV sample with a theoretical $\mu=22+\mu_1\cdot {\mathbf t}$ ($\mu_1=-0.5,-0.1,0.1\text{ or }0.5$, where t stands for the time) trend in the location parameter, while for the scale and shape $\sigma=10,-0.5\le\xi\le 0.5$ holds. The choice of the parameters was motivated by the average of MLEs of GEV paramaters for our real life data set. The generated sample size was varying between $20$ and $80$. The simulated power of the $LR_1$ and $LR_2$ tests (Table \ref{table:power}) show three important phenomena:
\begin{enumerate}
	\item For stronger trend (bigger absolute value of $\mu_1$), less observation is sufficient.
	\item Higher absolute value of tail index makes the detection easier (smaller sample size is acceptable).  By \cite{trend} this occurs as holding the scale parameter fixed results in observed data that are mode clustered around the location ($\mu$) for increasing $|\xi|$, than for xi=0 thereby allowing more precise estimates of the location coefficients.
	\item The direction of effect on location (sign of $\mu_1$) is not important.
\end{enumerate}
For obtaining $80\%$ power with high tail index ($\xi=0.5$) at least $30$ years of observation are needed in case of $\mu_1=0.5$. Instead, if $\mu_1=0.1$ than around $80$ years of observations necessary. In contrast, if the tail index is small ($\xi\sim 0$), for an adequate power $40$ years of observation are needed in case of strong trend ($0.5$), but more than $90$, if the trend is small. This result coincides with the fact, that only stations, with a trend coefficient larger than $0.35$ were detected as significant by the bootstrap simulations. 
Our observation is also in accordance with the conclusions in \cite{hasonlo}.

We followed the analysis by calculating the power of $LR_3$ and $LR_4$ tests. We simulated samples from GEV distribution with $\mu=0, \sigma = 10$ and  different shape parameters $-0.5\le\xi\le0.5$. Then we added the obtained GEV sample with the linear trend $22+\mu_1 {\mathbf t},$ with $-0.5\le \mu_1 \le 0.5$. The considered sample sizes varied between $20$ and $80$.
Based on $1000$ simulations we calculated the ratio of rejecting the null hypothesis using the $LR_3$ and $LR_4$ tests and the results are presented in Table \ref{table:power}. 

The results show that for $LR_3$ and $LR_4$:
\begin{enumerate}
	\item For stronger trend, less observation is sufficient.
	\item Lower value of tail index makes the detection easier (smaller sample size is sufficient).
	\item The direction of effect on location is not important.
	\item Theil-Sen estimator is slightly less sensitive for high tail index than least square method
\end{enumerate}

A big difference from the test described in Section \ref{GEV_paratrend} is that the tail index ($\xi$) has an important role. Higher shape parameter results in more extreme values - which behaves as an outlier in linear regression - and might have an unwanted effect on the parameter estimates.
%Connection with residuals problem? i.e. is there a higher autocorrelation for higher xi?
To sidestep this problem one can use robust estimations instead of classical linear model. Using Theil-Sen estimates, we could receive slightly better results. 

\begin{figure}[h!]
	\centering
	\includegraphics[width=0.8\textwidth]{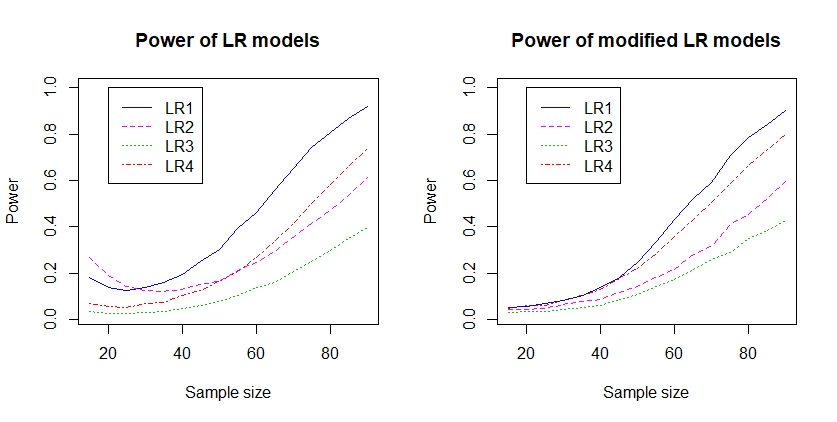}
	\caption{Comparison of power of $LR_1$, $LR_2$, $LR_3$ and $LR_4$ tests based on $10000$ simulations (left). Modified $LR$ tests are also included (right). Data was simulated from $GEV(22+{\mathbf t}\cdot 0.1,10,0.5)$ distribution of sizes between $15$ and $90$. }
	\label{fig:power}
\end{figure}

In Figure \ref{fig:power} one can compare the power of $LR_1$, $LR_2$, $LR_3$ and $LR_4$ tests on $GEV(22+{\mathbf t}\cdot c,10,0.5)$ data $(c=-0.5,-0.1,0.1\text{ or }0.5)$. The simulation shows that the power of $LR_1$ is the highest, followed by $LR_3$ and $LR_4$ is the worst as expected. Type $1$ error was highly different under $LR_1$ and $LR_4$ model, thus setting it to $0.05$ by using different critical values (previously calculated by $10000$ simulation) on likelihood ratio test could result in a more objective comparison between the models. By Figure \ref{fig:power} one can say, that $LR_1$ is still better using the corrected versions, however modified $LR_4$ model is also competitive and can be useful in some cases.

We highlighted the case of $\xi=0.5$, where the difference is stronger, similar result can be seen in other cases too, but with smaller difference.

\begin{table}[h]
	%\footnotesize
	%\small
	\centering
	\caption{Simulated power of the likelihood ratio tests of the trend under $LR_1$,$LR_2$, $LR_3$ and $LR_4$ calculated by $1000$ simulations from $GEV(22 +{\mathbf t} \cdot \mu_1,10,\xi)$ distribution with varying shape parameters and different sample sizes.}
	\label{table:power}
	\begin{tabular}{||cc|c|c|c|c|c|c|c|c|c|c|c|c||   }
		\hline\hline
		&&\multicolumn{3}{|c|}{$LR_1$}&\multicolumn{3}{|c|}{$LR_2$}&\multicolumn{3}{|c|}{$LR_3$}&\multicolumn{3}{|c||}{$LR_4$}\\
		\multicolumn{2}{||c|}{Sample size:} & 20&40&80& 20&40&80& 20&40&80& 20&40&80 \\
		$\mu_1$&$\xi$&&&&&&&&&&&&\\
		\hline\hline
		-0.5&$-0.5$ &0.516&0.996&1&0.216&0.9&1&0.397&0.981&1 &0.345&0.981&1\\
		\hline
		-0.5&$0$&0.248&0.930&1&0.232&0.918&1&0.244&0.901&1&0.241&0.899&1\\
		\hline
		-0.5&$0.5$&0.410&0.982&1&0.298&0.878&0.978&0.125&0.648&0.952&0.247&0.932&1\\
		\hline
		-0.1&$-0.5$&0.240&0.196&0.780&0.114&0.082&0.524&0.117&0.169&0.749&0.101&0.145&0.679\\
		\hline
		-0.1&$0$&0.126&0.128&0.540&0.142&0.122&0.496&0.068&0.092&0.467&0.07&0.091&0.5\\
		\hline
		-0.1&$0.5$&0.108&0.202&0.786&0.168&0.098&0.454&0.031&0.043&0.289&0.064&0.094&0.609\\
		\hline
		0.1&$-0.5$&0.224&0.228&0.806&0.122&0.086&0.546&0.106&0.163&0.731&0.126&0.142&0.709\\
		\hline
		0.1&$0$&0.13&0.11&0.530&0.12&0.114&0.482&0.064&0.097&0.499&0.086&0.089&0.470\\
		\hline
		0.1&$0.5$&0.128&0.214&0.792&0.172&0.128&0.426&0.035&0.044&0.308&0.053&0.096&0.575\\
		\hline	
		0.5&$-0.5$ &534&0.992&1&0.262&0.874&1&0.341&0.985&1&0.332&0.987&1 \\
		\hline
		0.5&$0$&0.286&0.910&1&0.266&0.908&1&0.235&0.914&1&0.226&0.881&1\\
		\hline
		0.5&$0.5$&0.432&0.984&1&0.3&0.862&0.974&0.139&0.647&0.952&0.211&0.931&1\\
		\hline\hline
		
	\end{tabular}
\end{table}

\subsection{Conclusion of simulations}\label{simres}

After numerous simulations one can say, that around $30-40$ size sample is required for $LR_1$ and $LR_2$ tests to reduce the type 1 error probability below $0.01$, in contrast with the regression type models with $LR_3$ and $LR_4$ tests, where $20$ observations might be enough. However, with modified critical values for $LR$ tests, smaller sample analysis become also reliable.

We observed, that $LR_1$ test is the most powerful test among the analysed ones. Regression based models ($LR_3, LR_4$) can not reach its level, however the power by robust regression with Theil-Sen estimation ($LR_4$) approaches the power of $LR_1$. The $LR_2$ test might be useful for larger size samples - for $n<100$ the possible trend in variance can disturb the trend estimation, which makes this model less powerful. In practice it can be useful, to analyse the data with almost similarly powerful, but different test. If one of the tests rejects to $H_0$ hypothesis, that means, there is a significant trend in the sample under one of the models. It can ba a potential indicator of stations, where further investigation is necessary.

Another concern is the different type 1 error for $LR$ tests, which could make the power incomparable. Solving this problem, we simulated critical values for each test, to set type 1 error to $0.05$ and the result was similar, still $LR_1$ test is the best under the assumption that there is not a true trend in the scale parameter of GEV.

Defining the critical values for modified LR tests in general case is not possible, since the parameters of the $H_0$ hypothesis and even the sample size has an effect on it. One need to simulate the critical values for each of the applied test and datasets separately.

\section{Analysis on real data}\label{real}

Now we use the simulation results of the trend detection methods on the Canadian fire weather index dataset \cite{fwidata}. First, we apply the methods on the data and calculate the estimated trend and p-value using the $LR$ tests without modification (Table \ref{table:data}). Significant, or potentially significant ($p<0.1$) trend appears at the following stations: $68,105,363,365,401,427,447,791$. On the following, we will analyse only these stations observations. We use bootstrap simulations for testing the regularity of samples from each stations and for constructing confidence interval of the estimated trend. Finally, we present the results on the 5 years return level, which is obtained both analytically and by simulation.

%Q: include modified values
% No, only for LR_1, next table
\begin{table}[h!]
	\centering
	\caption{General results for the analysed dataset, including station id, observation number, estimated trend in $\mu$ under different models and p-values of the given likelihood ratio test. We noted with * the ones, which were considered significant (or almost significant: $p<0.1$ based on at least one of the $LR_1$,\dots, $LR_4$ tests.) by a model.}
	\label{table:data}
	\begin{tabular}{||c|c|c|c|c|c|c|c|c|c||   }
		\hline\hline
		&&\multicolumn{4}{|c|}{Estimation of trend $mu_1$}&\multicolumn{4}{|c||}{p value}\\
		Station ID &Size&$M1.mu$&$M2$&$M3$&$M4$&$LR_1$ &$LR_2$&$LR_3$&$LR_4$ \\
		\hline\hline
		19&45&0.13&0.15&0.13&0.13&0.19&0.21&0.19&0.2\\
		\hline
		68 *& 17&-0.27&-0.46&-0.48&-0.54&0.43&\textbf{0.000}&0.54&0.64 \\
		\hline
		75&35&0.02&0.00&-0.05&-0.00&0.81&0.99&1&1\\
		\hline
		105 *&40&0.22&0.17&0.13&0.17&\textbf{0.006}&\textbf{0.04}&\textbf{0.01}&\textbf{0.008}\\
		\hline
		145&41&-0.04&-0.02&-0.03&-0.07&0.81&0.92&0.81&0.91\\
		\hline
		161&45&0.08&0.09&0.11&0.14&0.45&0.4&0.48&0.6\\
		\hline
		363 *&19&0.22&0.54&0.44&0.49&0.49&\textbf{0.08}&0.75&1\\
		\hline
		365 *&14&-1.01&0.94&-0.51&-0.33&\textbf{0.04}&\textbf{0.000}&0.18&0.28\\
		\hline
		401 *&29&-0.11&-0.35&-0.07&-0.06&0.53&\textbf{0.09}&0.57&0.58\\
		\hline
		427 *&45&-0.23&-0.23&-0.25&-0.23&\textbf{0.01}&\textbf{0.01}&\textbf{0.01}&\textbf{0.01}\\
		\hline
		447 *&19&0.54&0.48&0.27&0.4&0.178&\textbf{0.07}&0.26&0.2\\
		\hline
		791 *&21&-0.74&-0.62&-0.185&-0.55&\textbf{0.0008}&\textbf{0.001}&1&\textbf{0.002}\\
		\hline
		865&11&0.07&-0.13&-0.16&0.11&0.94&0.91&1&0.95\\		
		\hline\hline
		
	\end{tabular}
\end{table}

A strange phenomena of Table $\ref{table:data}$ is the likelihood ratio tests results on observations from station $791$. Three of the tests detect significant trend, while $LR_3$ test has p-value of $1$. This is caused by a critical extreme value, which can be handled by using model M1.mu, M2. Even model M4 using Theil-Sen regression is not sensitive for a single value, however for classical linear regression in model M3 an outlier can ruin the effectiveness.

In further analysis we will mainly use $LR_1$ test as the most powerful one, thus we present the simulated critical values under $LR_1$ test using observations from each stations with potential significant trend. Additionally, we calculated the p-values for modified $LR_1$ tests (Table \ref{table:critval}).

\begin{table}[h!]
	\centering
	\caption{Simulated critical values for using modified $LR_1$ test. Values were calculated after $10000$ simulations}
	\label{table:critval}
	\begin{tabular}{||c|c|c|c|c|c|c|c|c||   }
		\hline\hline
		Station ID & 68&105&363&365&401&427&447&791\\
		\hline\hline
		Critical value & 6.399&4.314&6.657&11.4&5.84&4.243&8.831&5.314 \\
		\hline
		P-value & 0.547&0.012&0.552&0.211&0.618&0.013&0.322&0.005 \\
		\hline\hline
		
	\end{tabular}
\end{table}

\subsection{Bootstrap resampling from real life data }

Using the different methods on the dataset can provide us trend estimations, however the special structure of extreme value data can easily lead to false significant detection or non-detection. Therefore we use bootstrap simulations to estimate the possible error in each cases.

From the real dataset, we drew independent bootstrap samples \cite{efron} (without replacement) of the size of the original data for each station (permutation of the sample). The sample elements are chosen independently, thus no trend should be detected. Repeating the sampling $5000$ times and using the same likelihood ratio test at nominal significance level $0.05$, we experienced, that for the stations, where trend was detected the ratio of falsely discovered trend under both $LR_1$ and $LR_2$ test was high (see first two lines of Table \ref{table:realdata}). The cases in Table \ref{table:realdata} are all type $1$ errors, which should be around $5\%$ for an adequate statistical test. Presented high percentages might be the result of some outliers on the beginning or in the end of the test period, which can be corrected only by longer test periods (i.e. larger sample size).
A similar analysis was executed, by using $50$ size bootstrap samples with replacement to extend the theoretical test period. After $5000$ repetitions, the type $1$ error were clearly lower than in the previous case (see the third and fourth row of Table \ref{table:realdata}). We did not use modified $LR$ tests here, since those tests were constructed to set type 1 error probability to $0.05$.

\begin{table}[h!]
	\centering
	\caption{Bootstrapped type 1 error of the LR tests of the trend in location parameter of GEV at the significance level 0.05. Tests were conducted for independent permutations of the original data.
		Total number of simulations was $5000$ for each cases.}
	\label{table:realdata}
	\begin{tabular}{||c|c|c|c|c|c|c|c|c||   }
		\hline\hline
		Station ID & 68&105&363&365&401&427&447&791\\
		\hline\hline
		Type 1 error: real sample size ($LR_2$) & 0.146&0.064&0.134&0.282&0.089&0.061&0.109&0.168 \\
		\hline
		Type 1 error: real sample size ($LR_1$) & 0.097&0.065&0.123&0.163&0.096&0.064&0.126&0.33 \\
		%\hline
		%Type 1 error: real sample size (modified $LR_1$) & 0.014&0.044&0.017&0.001&0.024&0.053&0.002&0.16 \\
		\hline
		Type 1 error: sample size 50 ($LR_2$) & 0.062&0.06&0.061&0.069&0.073&0.062&0.058&0.05 \\
		\hline
		Type 1 error: sample size 50 ($LR_1$)& 0.049&0.068&0.066&0.074&0.079&0.063&0.066&0.235 \\
		\hline\hline
		
	\end{tabular}
\end{table}
%Q: similar for modified
% Strange result, theoretically should be around 0.05, but if the distribution does not fit, than anything can be

As can be seen in Table \ref{table:data}, $6$ of the stations could satisfy the size condition that was mentioned in Section \ref{simres} (at least $30$ size for $LR_1$), thus only stations $105$ and $427$ show significant difference from iid. GEV. In observations of stations $365$ and $791$ significant trend was also detected, however the sample size is less than $20$. Nevertheless, even with high type $1$ error, $LR_1$ test could achieve significance detection with high probability even for smaller samples if there is a large existing trends.

The results of Tables \ref{table:data} and \ref{table:critval} suggest that at the significance level 0.05 for stations $105$, $427$ and $791$ there is a strong trend in fire weather index using modified $LR_1$ test, moreover for stations $68$, $363$, $365$, $401$ and $447$ there is a chance that they have a trend (significant using $LR_1$ or $LR_2$). To test the effectiveness of the likelihood ratio test, we constructed a bootstrap simulation based method. In contrast to \cite{trend}, we do not investigate trend in shape (M2 model), only in the location parameter (M1.mu model), due to the small sample size.

For each station, we estimate the parameters of either M1.mu or M2 and we calculate the estimated distribution of each sample element and apply the transformation to the standard Gumbel distribution based on this distributions (equation \ref{stgumbel}). Theoretically we receive a standard Gumbel distributed sample. Taking a bootstrap sample from it, and transforming it back can result in a time series with similar expected properties, which is suitable for using the likelihood ratio test on it.

From the transformed standard Gumbel-distributed observations we took samples of the same size as the original time series, without replacement (shuffle). After the inverse transformation we expected to receive also a significant trend in the location parameter. We simulated $1000$ different bootstrap samples without replacement for the original size analysis and calculated if there was a detected trend using the likelihood ratio test (statistical power). Additionally, we simulated $50$ size data from $GEV(0,1,0)$ and transformed it back to original distribution using the previously estimated parameters. This method coincides with the parametric bootstrap based method. The power of $LR$ test at analysed stations can be seen in Table \ref{table:real_param}. We also used modified $LR_1$ test for power calculation. 

%At the first sight, we experienced such strange results, that for higher bootstrap sample the simulated power were smaller in many cases. We believe, this was caused by the fact, that $LR_1$ and $LR_2$ test statistics are asymptotically $\chi^2$ distributed and the simulated samples usually have small size. Therefore we approximated the critical values of the $LR$ for small size samples by $10000$ simulations. In the real life calculations we used these simulated critical values for $LR$ test decisions.
%Q: Ezt kellene az egész fejezetben?
% Úgy érzem megvan

\begin{table}[h!]
	\centering
	\caption{Power calculation by bootstrap simulations using $LR_1$, $LR_2$ and modified $LR_1$ tests. For $LR_1$ and $LR_2$ we also took $50$ size bootstrap samples. Total number of simulations was $1000$ in each case.}
	\label{table:real_param}
	\begin{tabular}{||c|c|c|c|c|c|c|c|c||   }
		\hline\hline
		Station ID & 68&105&363&365&401&427&447&791\\
		\hline\hline
		Power: real sample size ($LR_2$) & 0.803&0.515&0.206& 0.283 &0.521&0.718&0.324&0.768 \\
		\hline
		Power: real sample size ($LR_1$) & 0.312&0.415&0.197& 0.028 &0.383&0.750&0.244&0.971 \\
		\hline
		Power: real sample size (modified $LR_1$) & 0.087&0.574&0.069& 0.489 &0.081&0.730&0.241&0.892 \\
		\hline
		Power:  sample size 50 ($LR_2$) &  0.943 &0.733&0.770& 0.533 &0.921&0.847&0.985&0.866 \\
		\hline
		Power: sample size 50  ($LR_1$) &  0.998  &0.495&0.354& 0.717 &0.573&0.902&0.986&1 \\
		\hline\hline
		
	\end{tabular}
\end{table}
% station 365: too small sample, critical values vere only calculated from 15 size samples. Manually modified

In both M1.mu and M2 models, we can observe, that the bootstrapped type 1 error is slightly higher than the ordinary ($0.05$) (Table \ref{table:realdata}). 
%Since modified tests has exactly $0.05$ type $1$ error, it is more effective to compare the power values using modified $LR_1$ and $LR_2$. 
We can observe, that the power is slightly less than or around the generally accepted ($0.8$) (Table \ref{table:real_param}). For higher power we receive higher type $1$ error too, thus the effectiveness of the likelihood ratio test can not reach the ideal, but is still promising. However, for larger bootstrap samples (of size 50 in our case), the type $1$ error converges to $ 0.05$, which suggests, that $50$ might be considered as a sample size, which is sufficient for the likelihood ratio test to work properly in detecting a linear trend in the location parameter of the GEV distribution.

\subsection{Confidence interval for trend on real life results}\label{CI_section}

Estimating trend in the GEV parameters might be an interesting problem to identify the properties of the stations. However, it is hard to use these informations directly. The realistic usage of the identified trend can be a prediction for the upcoming years. Unfortunately, a single value is not adequate to describe the risk in predictions. This motivated us to construct confidence intervals for the estimated trend on location parameter.

For constructing the confidence interval, we used parametric bootstrap simulations the same way, as it is described in the previous section. We estimated the parameters of the GEV model with trend in location parameter (M1.mu model) by maximum likelihood method. We transformed the data into a standard Gumbel distribution by using the MLE of the parameters as described in Equation \ref{stgumbel}. We took bootstrap samples from the transformed data and applied the inverse transformation. The obtained bootstrap sample had the same theoretical properties as the original data. We fitted the GEV model with trend in $\mu$ parameter to the bootstrap sample. By calculating the $0.025$ and $0.975$ quantiles of MLE of $\mu_1$ we could construct a two sided $95\%$ confidence interval. 

It is worth noting, that an additional goodness of fit test of the unit GEV distribution on the transformed data showed that stations $365$, $427$ and $791$ can not be considered as unit GEV distributed at the significance level 0.05 either using Anderson-Darling test (p values: $0.041, 0.021,5\cdot 10^{-5}$) or Cramér-von Misses test (p values: $0.035, 0.048,2\cdot 10^{-4}$). This might be a theoretical problem by using the method, however the speciality of data can also cause this phenomena.

In most  cases, the confidence interval covers $0$, thus we don't reject the hypothesis that there is no trend in the location parameter of the GEV at the significance level $0.05$. For predictions, the obtained confidence interval can still determine the uncertainty, but questions the usage of such a complex estimating procedure. However, for some stations, based on the confidence interval we concluded that there is a significant trend in the location parameter at the significance level $0.05$, see Table \ref{table:ci}. We also included confidence interval estimations using $M2$,$M3$ and $M4$ models (described in Section \ref{section_LM}). 

\begin{table}[h]
	\centering
	\caption{Confidence intervals for the trend coefficient under different models for stations with significant trend detected detected by at least one of the tests $LR_1,\dots,LR_4$ in Table \ref{table:data}. Results are based on $1000$ bootstrap simulations.}
	\label{table:ci}
	\begin{tabular}{||c|c|c|c|c||   }
		\hline\hline
		Station & M1.mu &M2&M3& M4\\
		\hline\hline
		68&-1.24 ; 0.26&\textbf{-0.9; -0.02} &-1.2 ; 0.23&-1.44 ; 0.26\\
		\hline
		105&\textbf{0.09 ; 0.35}&\textbf{0.01 ; 0.32}&-0.03 ; 0.3&\textbf{0.001 ; 0.34}\\
		\hline
		363&-0.57 ; 0.97&-0.26 ; 1.14&-0.28 ; 1.12&-0.32 ; 1.27\\
		\hline
		365&\textbf{-1.88 ; 0}&-0.128 ; 0.28&-1.3 ; 0.24&-1.32 ; 0.6\\
		\hline
		401 &-0.76 ; 0.17&-0.67 ; 0.03&-0.39 ; 0.24&-0.45 ; 0.32\\
		\hline
		427&\textbf{-0.47 ; -0.06}&\textbf{-0.4 ; -0.07}&\textbf{-0.42 ; -0.08}&\textbf{-0.44 ; -0.04}\\
		\hline
		447&-0.02 ; 1.1&-0.26 ; 0.81&-0.21 ; 0.86&-0.22 ; 1.03\\
		\hline
		791&\textbf{-0.98 ; -0.34}&\textbf{-0.96 ; -0.23}&\textbf{-3.36 ; -0.38}&\textbf{-1.06 ; 0}\\
		\hline\hline		
	\end{tabular}
\end{table}

\subsection{Final results}

Using the results of previous sections we could claim, that a time series can be modelled by a GEV model with nonzero trend $\mu_1$ if satisfies the following conditions in M1.mu:
\begin{itemize}
	\item At least $20$ size sample, to approach the $0.05$ probability of Type 1 error of the test (Table \ref{table:type1}, Figure \ref{fig:type1})
	\item Reaching usual $0.8$ power, calculated from the sample by bootstrap simulations or using theoretical power based on the estimated parameters (Table \ref{table:power})
	\item Significant trend must be detected under $LR_1$ test (or best by its modification at the small sample sizes)
\end{itemize}
%More comments

In Table \ref{table:stations} we collected the most important results of the previously promising stations under $LR_1$ test and its modified version, as it was declared as the most powerful test. Station $105$, $427$, and probably $791$ can satisfy all of the conditions using original $LR_1$ test, thus only in these stations the trend can be confirmed. Station $365$ might also be promising by $LR_1$, because the p-value is small and the power is large enough, but the sample size is dangerously small, which can mislead us. We suggest an additional $5-10$ years of observations for an appropriate analysis. In station $401$ the sample size is acceptable, however the power and the significance level are not adequate. We believe this is caused by the $-0.12$ trend, which is relatively small and might require larger sample size for significant detection. In the other two stations, we can not confirm a trend in location parameter of GEV. Modified $LR_1$ test supports stations $105$, $427$ and $791$ as significant ones while oppose $365$ and $401$. Since modified $LR_1$ was considered the most powerful test, we must accept the three stations with significant detected trend.

The confidence intervals for parameter $\mu_1$ (see Table \ref{table:ci}) also approve these observations. Moreover, under linear regression with GEV residuals  stations $105,427$ and $791$ were also marked as those with a significant trend.

\begin{table}[h]
	\centering
	\caption{Detailed results for every stations with potential significant trend using M1.mu model. We present the p-value and power of $LR_1$ test, type 1 error calculated by bootstrap simulations and power of modified $LR_1$ test.}
	\label{table:stations}
	\begin{tabular}{||c|c|c|c|c|c|c|c|c|c||   }
		\hline\hline
		Station&Size &$\hat{\mu_0}$&$\hat{\mu_1}$&$\hat{\sigma}$&$\hat{\xi}$&$LR_1$&Modified $LR_1$&Power&Modified power\\
		\hline\hline
		68&17&18.01&-0.27&8.18& -0.09&0.43& 0.547& 0.431&0.087\\
		\hline
		105&40&10.53&0.22&5.77&0.01& 0.006&0.012&0.438&0.574\\
		\hline
		363&19&14.29&0.22&7.42&-0.09&0.489&0.552&0.422&0.069\\
		\hline
		365&14&23.92&-1&8.2&-1&0.038&0.211&0.654&0.489\\
		\hline
		401&29&19.27&-0.12&7.06&-0.21&0.096&0.618&0.428&0.081\\
		\hline
		427&45&22.3&-0.23&7.29&-0.21&0.001&0.013&0.787&0.730\\
		\hline
		447&19&19.68&0.54&5.53&-0.02&0.177&0.322&0.515&0.241\\
		\hline
		791&21&33.78&-0.74&6.03&0.59&0.0007&0.005&0.980&0.892 \\
		\hline\hline
		
	\end{tabular}
\end{table}

Practical benefit of identifying a trend in data at the stations is to predict upcoming values. However, association with GEV distribution result in big variance, thus single predicted values can not be able to describe the upcoming events properly. Beside confidence intervals - that we calculated in previous sections - return level is a useful value for measuring the expected risk. Return level of $k$ year gives a value, which will be reached in the next $k$ years with $50\%$ chance. In case of iid. sample it is easy to calculate, but with significant trend in the location parameter the method is not straightforward. The different distributions of the independent variables cause that in the beginning or in the end of the period there is a higher chance to reach the return level (depending on the direction of trend). Calculating the return levels in an analytic way is problematic in this case, because the uncertainty of parameters are high. Thus we used a bootstrap simulation based method for estimating return levels and confidence intervals. 
Another solution of this problem can be found in \cite{ribei}.

With $\mu_0, \mu_1, \sigma, \xi$ parameters, after ${\mathbf t}$ years the distribution of monthly maxima can be approximated by $GEV(\mu_0+\mu_1 {\mathbf t}, \sigma,\xi)$. The expectation of the upcoming $k$ years maxima can be received by solving

\begin{eqnarray}\label{returnlevel}
	P\big(\max_{j=1}^{k} X_k<y\big)&=&P(X_1,X_2,\dots, X_k<y)=P(X_1<y)\cdot P(X_2<y)\cdot\dots\cdot P(X_k<y)=\nonumber\\
	&=&\prod_{j = 1}^{k}\exp\bigg(\big(1-\xi\big(\frac{y-\mu_0 -j\cdot \mu_1}{\sigma}\big)\big)^{-1/\xi}\bigg)=\frac{1}{2}
\end{eqnarray}

where $y$ is the desired $k$ years return level. In our estimation we set $\mu_0$ as the estimated locations parameter of the last observation, while $\mu_1$ is the trend coefficient by model M1.mu. For a given model, one can solve the equation (\ref{returnlevel}) numerically.

\begin{figure}[h!]
	\centering
	\includegraphics[width=0.8\textwidth]{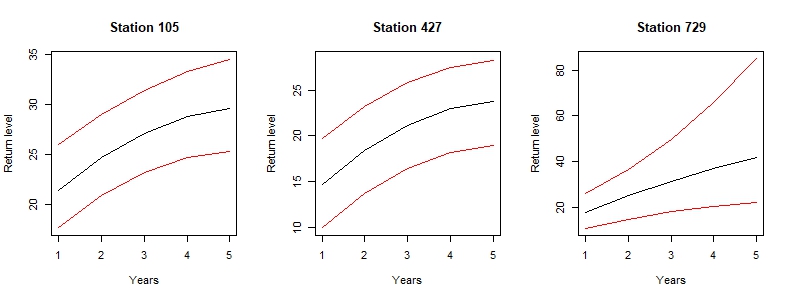}
	\caption{Return level and confidence interval for station $105,427$ and $791$.}
	\label{fig:returnlevel}
\end{figure}

To include uncertainty of parameters to return level estimation, we used bootstrap samples for parameter estimation applying transformation described in equation (\ref{stgumbel}). For each bootstrap sample we were able to calculate a return level for a $5$ years long period. The mean of the calculated return levels approximates the true value well, while with $0.025$ and $0.975$ quantiles the $95\%$ confidence interval can be estimated. We used $1000$ simulations for each of the significant stations. In Figure \ref{fig:returnlevel} one can see, the estimated values of return level and confidence interval.

\section{Conclusion}

We analysed the effectiveness of several likelihood ratio tests for extreme values to use it in real life - forest fire weather index related - problems. We presented four tests for testing trend, one based on GEV distribution with trend in parameters, the other fits linear regression but assuming the residuals as GEV. 

The first model uses likelihood ratio to identify significant trend, based on two approaches: one is an extreme value model including possible trend in location and scale parameter and the other can contain trend in the location parameter only. Our bootstrap resampling simulations using the available data showed that the likelihood ratio test has larger type I. error than accepted in the real life cases. Simulations from theoretical distributions resulted in suggestions for the sufficient sample size for acceptable type I. error (0.1), which is around $25-30$. On the other hand, for existing trend the required sample size for true detection depends on the margin of trend. Generally around $30$ sample  size is enough for regular cases. With parametric bootstrap simulations, we were also able to construct confidence intervals for the estimated trends.

The second model uses least squares method to estimate trend and fits GEV for the residuals. We also use likelihood ratio test between the regression model and an iid. GEV sample, to identify significant difference. With bootstrap simulations we constructed confidence intervals for estimated trends. Moreover, with simulated time series we proved remarkable dependence on GEV's shape parameter. In optimal cases, we could show that around $30$ size samples could be sufficient for detecting trend with relevant margin.

We applied power analysis under both model, where the likelihood ratio test was modified, in order to set type $1$ error to $0.05$. The test distribution was $GEV(22+0.1\cdot t,10,0.5)$. The observed power was higher by using $LR_1$ model, thus for heavy tailed data using $LR_1$ model is suggested over $LR_3$ and $LR_4$ models. The $LR_2$ method only suggested when significant trend assumed in scale parameter, otherwise much more type $1$ error and unpredicted bias appears.

Using these observations we re-evaluated the likelihood estimation on the data. We only accepted the result where the size was at least $20$ (slightly less than the required, but still can be representative). From the original $13$ time series, 3 stations still satisfy the conditions and contain significantly detectable trend under both approaches.

\section{Acknowledgement}

This research was supported by the European Union, co-financed by the European Social Fund (EFOP-3.6.3-VEKOP-16-2017-00002).

\end{document}